\documentclass[11pt]{article}
\usepackage{fullpage}
\usepackage{latexsym,amsmath,amssymb}
\usepackage{graphicx,subfigure}
\usepackage[ruled,vlined]{algorithm2e}
\usepackage{enumerate}
\usepackage{enumitem}
\usepackage{color}

\newtheorem{definition}{Definition}
\newtheorem{lemma}{Lemma}
\newtheorem{corollary}{Corollary}
\newtheorem{theorem}{Theorem}
\newtheorem{example}{Example}
\newtheorem{observation}{Observation}
\newenvironment{pf}{\begin{trivlist}
\item[\hspace{\labelsep}{\em\noindent Proof: }]
}{\hfill$\Box$\end{trivlist}}

\begin{document}

\title{On a conjecture of compatibility of multi-states characters}

\author{Michel Habib\thanks{Universit\'e Paris Diderot - Paris 7, LIAFA, Case 7014,
    75205 Paris Cedex 13, France.  E-mail:  michel.habib/thu-hien.to/@liafa.jussieu.fr.}~,~~Thu-Hien To$^*$}

\maketitle

\begin{abstract}
Perfect phylogeny consisting of determining the compatibility of a set of characters is known to be NP-complete \cite{BFW92,S92}. We propose in this article a conjecture on the necessary and sufficient conditions of compatibility: Given a set $\mathcal{C}$ of $r$-states full characters, there exists a function $f(r)$ such that $\mathcal{C}$ is compatible iff every set of $f(r)$ characters of $\mathcal{C}$ is compatible. According to \cite{F75,EJM76,F77,M83,G91,LGS09}, $f(2)=2$, $f(3)=3$ and $f(r) \ge r-1$. \cite{LGS09} conjectured that $f(r) = r$ for any $r \ge 2$. In this paper, we present an example showing that $f(4) \ge 5$ and then a closure operation for chordal sandwich graphs. The later problem is a common approach of perfect phylogeny. This operation can be the first step to simplify the problem before solving some particular cases $f(4), f(5), \dots$, and determining the function $f(r)$.
\end{abstract}


\section{Introduction}
Given an input biological data of a currently-living species set, phylogenetics aims to reconstruct evolutionary history of their ancestors. The evolutionary model of perfect phylogeny is phylogenetic tree, and the data are characters of species. Characters can be morphological, biochemical, physiological, behavioural, embryological, or genetic. Each character has several states. Here are some examples. The character \textit{have wings} has two states: with wings and without wings. The character \textit{number of legs} has many states: one leg, two legs, four legs, ... These are morphological characters. For an example of genetic characters, given a set of DNA sequences having a same length, if we consider each position on the sequences to be a character, then each character has $4$ states corresponding to $4$ bases of DNA as \textit{A, T, C, G}. 

Let $\mathcal{L}$ be a species set, and let $c$ be a character on $\mathcal{L}$. Then, $c$ can be represented by a partition of a non-empty subset $\mathcal{L}'$ of $\mathcal{L}$ such that each part consists of all species having the same state of $c$. So a set of characters is a set of partitions.
A character is said to be trivial if the partition has at most one part having more than $1$ element. Otherwise, it is non-trivial. 
If $\mathcal{L}'=\mathcal{L}$, then $c$ is a full character, otherwise it is a partial character. If $c$ has at most $r$ parts, then $c$ is a $r$-states character. A binary character is a $2$-states full character. 

\begin{example}
\label{ex:character1}
Let $\mathcal{L}=\{a,b,c,d,e,f,g\}$ and $\mathcal{C}=\{c_1,c_2,c_3,c_4\}$ be a collection of characters on $\mathcal{L}$ such that:

$c_1 = ab|cdefj|ghi$, $c_2 = def|abcghij$, $c_3 = gh|defi$, $c_4 = abcd|ghi$.

It means that $c_1$ has $3$ states $0,1,2$ such that the species $a,b$ are in state $0$, the species $c,d,e,f,j$ are in state $1$, and the species $g,h,i$ are in state $2$. Similarly for $c_2,c_3,c_4$.
\end{example}

A phylogenetic tree on a species set $\mathcal{L}$ is a tree where each leaf is labelled distinctly by a species of $\mathcal{L}$.

\begin{definition} \cite{DS92}
Let $c$ be a $r$-states character  and let $T$ be a phylogenetic tree on $\mathcal{L}$. 
For $i = 0, \dots, r-1$, denote by $T_i(c)$ the minimal subtree of $T$ on the leaf set consisting of the species having the state $i$ of $c$. So, $c$ is said to be \textbf{convex} on $T$ iff the subtrees $T_i(c)$ are pairwise vertex-disjoint.
\end{definition}

A set of characters is \textbf{compatible} iff there exists a phylogenetic tree on which every character is convex. So the problem of perfect phylogeny is also commonly known as the character compatibility problem.
\begin{example}
\label{ex:convex}The set of characters $\mathcal{C}$ in Example \ref{ex:character1} are compatible because there is a phylogenetic tree $T$ in Figure \ref{fig:PP} on which every character is convex. For example, $c_1$ is convex on this tree because the subtrees of $T$ on $\{a,b\}$, $\{c,d,e,f,j\}$ and $\{g,h,i\}$ are pairwise vertex-disjoint. Similarly, $c_2$, $c_3$, and $c_4$ are also convex of this tree.

\begin{figure}[ht]
\begin{center}
\includegraphics[scale=.7]{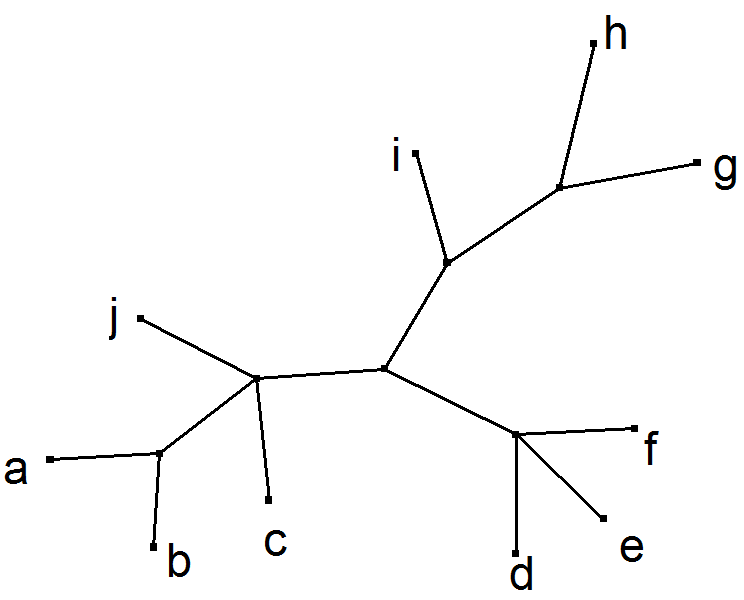}
\caption{A phylogenetic tree on which all characters in Example \ref{ex:convex} are convex}
\label{fig:PP}
\end{center}
\end{figure}
\end{example}

Determining the compatibility of a set of characters is NP-complete \cite{BFW92,S92}. In this article, we are interested in the necessary and sufficient conditions of compatibility of a set of $r$-states full characters. We propose the following conjecture.

\vspace{0.5cm}

\textbf{Conjecture:}
For any set $\mathcal{C}$ of $n$ $r$-states full characters, there exists a function $f(r)$, which does not depend on $n$, such that $\mathcal{C}$ is compatible iff every set of $f(r)$ characters of $\mathcal{C}$ is compatible.

\vspace{0.5cm}

This conjecture is based on the following previous results. According to \cite{EJM76,M83,G91}, $f(2)=2$ and according to \cite{LGS09}, $f(3)=3$.
Fitch-Meacham examples \cite{F75,F77,M83,LGS09} showed that $f(r) \ge r$ for any $r \ge 2$. 
There are polynomial algorithms of checking compatibility of $3$-states full characters \cite{DS92} and $4$-states full characters \cite{KW94}. By \cite{McMWW93}, if the number of states is restricted, then checking the compatibility of $\mathcal{C}$ is polynomial. In general, \cite{AFB94,KW95,AFB96} showed that there is a polynomial algorithm in the number of characters and species, but exponential in the number of states. 

\subsection*{Some related work}

\hspace*{.6cm}\textit{Existence of perfect phylogeny:} Given a set of characters on $\mathcal{L}$, is there any phylogenetic tree on which all characters are convex? It is easy to determine whether a collection of binary characters is compatible.  However the problem is NP-complete even for $2$-states characters \cite{BFW92,S92}. 
There exists effective, practical approaches for $2$-states characters \cite{GFB07}, and a new approach for this problem is proposed in \cite{G09}. 


\textit{Quartet problem:} A minimal non-trivial character is a $2$-states character such that each state contains exactly two species. Such character is called a \textit{quartet}. As stated in the previous paragraph, the problem of compatibility of a set of quartets is NP-complete \cite{BFW92,S92}. However, there are some particular cases that the problem is polynomial \cite{BDS99}, see \cite{SS03} for details.

\textit{Define a tree by characters:} a set of characters define a tree iff there is not any other tree on which these characters are convex. \cite{SS02} showed that a set of $3$ characters are not sufficient to define a tree but a set of $5$ characters are. Later, \cite{HMS05} showed that for any tree, there exist at most $4$ characters which define this tree. Hence, $4$ is the optimal value.
For the problem of whether a set of characters defines a tree, this is recently proved to be NP-hard \cite{HS10}. The author solved the quartet challenge proposed by \cite{SS03} as follows: given a phylogenetic tree on $\mathcal{L}$, and a quartet set $\mathcal{Q}$ which is convex on this tree. Is there any other tree on which this quartet set is also convex? Despite the NP-hardness, there is a polynomial algorithm for this problem when $|\mathcal{L}|-|\mathcal{Q}|=3$ \cite{BDS99,BBDS00}.

\textit{Maximum parsimony:} When there is no perfect phylogeny that can be inferred from data, it is desirable to find a model that minimize the number of reverse and convergent transitions. That is the problem of maximum parsimony.

\textit{Perfect phylogeny with recombination:} When the characters set are not tree-representable, it is also interesting to construct a model that can represent phylogenies. The model used here is recombination networks. Introduced by \cite{H90}, then intensive work have been done on this problem, including \cite{WZZ01,GEL03,GEL04A,GEL04B,G05,GBBS07}.


\section{Preliminaries}

A very popular approach of perfect phylogeny is using chordal completion of vertex-coloured graphs, or equivalently chordal sandwich graph. 

\begin{definition}
Let $\mathcal{L} = \{x_1, \dots , x_m\}$ be a species set and let $\mathcal{C}=\{c_1, \dots, c_m\}$ be a set of characters on $\mathcal{L}$. Each $c_i$ is a partition of a subset of $\mathcal{L}$. The \textbf{partition intersection graph} $G=(V,E)$ of $\mathcal{C}$ is constructed as follows:

- Each character of $\mathcal{C}$ is associated with a different colour.

- Each vertex of $V$ corresponds to a state of a character of $\mathcal{C}$. This vertex is then coloured by the colour of the character. 

- There is an edge between $2$ vertices if the $2$ corresponding states of the $2$ characters have at least a common species.
\end{definition}

In our figures, in stead of colouring the vertices, we include the name of the characters in the labels of the vertices.

\begin{example}\label{ex:Int_G}
Let consider the character set in Example \ref{ex:convex}: $\mathcal{C}=\{c_1,c_2,c_3,c_4\}$ where

$c_1 = ab|cdefj|ghi = c_{1,0}|c_{1,1}|c_{1,2}$, 

$c_2 = def|abcghij = c_{2,0}|c_{2,1}$, 

$c_3 = gh|defi = c_{3,0}|c_{3,1}$, 

$c_4 = abcd|ghi = c_{4_0}|c_{4,1}$. 

Each vertex $c_{i,j}$ represents the state $j$ of character $i$.
The partition intersection graph of $\mathcal{C}$ is in Figure \ref{fig:Int_G}.

\begin{figure}[ht]
\begin{center}
\includegraphics[scale=.7]{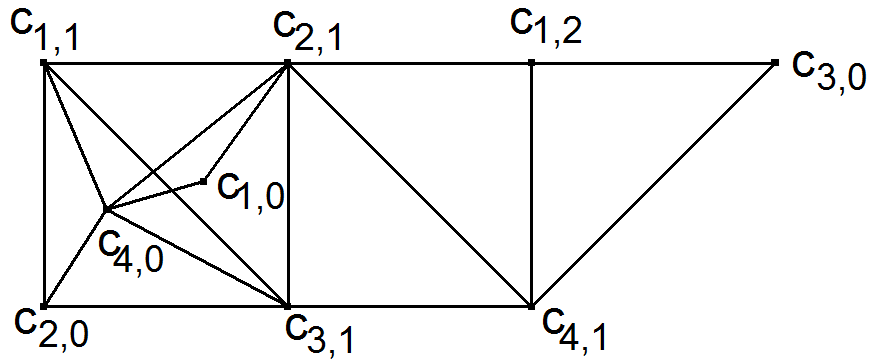}
\caption{An example of partition intersection graph}
\label{fig:Int_G}
\end{center}
\end{figure}

\end{example}

A graph $G$ is \textit{chordal} if every cycle of length $\ge 4$ contains at least a chord.
A \textit{chordal completion} of $G$ is a chordal graph $G'=(V,E')$ such that $E \subseteq E'$. This completion is minimal iff when we remove any edge in $E' \setminus E$, the resulting graph is not chordal. 

Given a vertex-coloured graph $G$, a \textbf{proper chordal completion} of $G$ is a chordal graph $G'=(V,E')$ such that $E \subseteq E'$ and $E'$ does not contain any edge connecting two vertices of the same colour.

\begin{theorem}\cite{B74,M83,S92}
A set of characters $\mathcal{C}$ is compatible iff its partition intersection graph has at least a proper chordal completion.
\end{theorem}

The set of characters in Example \ref{ex:Int_G} is compatible as showed in Example \ref{ex:convex}. Its partition intersection graph (Figure \ref{fig:Int_G}) has indeed a proper choral completion which is itself.

Proper chordal completion of vertex-coloured graph can be stated equivalently under the form of sandwich problems.

\begin{definition}
Given a graph $G = (V,E,F)$ where $F$ is a set of pairs of vertices of $G$ such that $E \cap F = \emptyset$. 

If there is a graph $G_S = (V,E_S)$ such that $E \subseteq E_S \subseteq E \times E \setminus F$ and $G_S$ satisfies property $\Pi$, then $G_S$ is called a $\Pi$\textbf{-sandwich graph} of $G$.
\end{definition}

See \cite{GKS95} for some problems and results on graph sandwich problems. 


It is easy to see that a chordal completion of a vertex-coloured graph $G=(V,E)$ is proper iff it is a chordal-sandwich graph of $G=(V,E,F)$ where $F$ is the set of pairs of vertices having a same colour. So, by considering the set $F$, we can ignore the colours of the initial graph.  We also call a \textit{chordal-sandwich graph} of $G$ a \textit{proper chordal completion} of $G$, i.e. a chordal completion of $G$ without using any pair of vertices in $F$.


\section{Our contributions}
Fitch-Meacham examples were first introduced in \cite{F75,F77}, then later generalized in \cite{M83} and formally proved in \cite{LGS09}, showed that $f(r) \ge r$ for any $r \ge 2$. \cite{LGS09} conjectured that for any $r$, there is a perfect phylogeny on $r$-state characters if and only if there is one for every subset of $r$ characters, i.e. $f(r) = r$ for any $r \ge 2$. However, we have an example in Section \ref{sec:example} showed that $f(4) \ge 5$.
It improves the lower bound of Fitch-Meacham examples and shows that the conjecture in \cite{LGS09} is not true. After that, in Section \ref{sec:partial}, we propose a closure chordal sandwich graph operation such that the obtained graph has a stronger structure. As a consequence, one can suppose that the input graph has such a structure, which can facilitate settling the conjecture.


\section{An example of $4$-States Characters}
\label{sec:example}
We present here an example of a set of $4$-states characters which is not compatible, but every $4$ characters of this set are compatible.

\begin{figure}[ht]
\begin{minipage}[b]{70mm}
\begin{center}
\includegraphics[scale=.75]{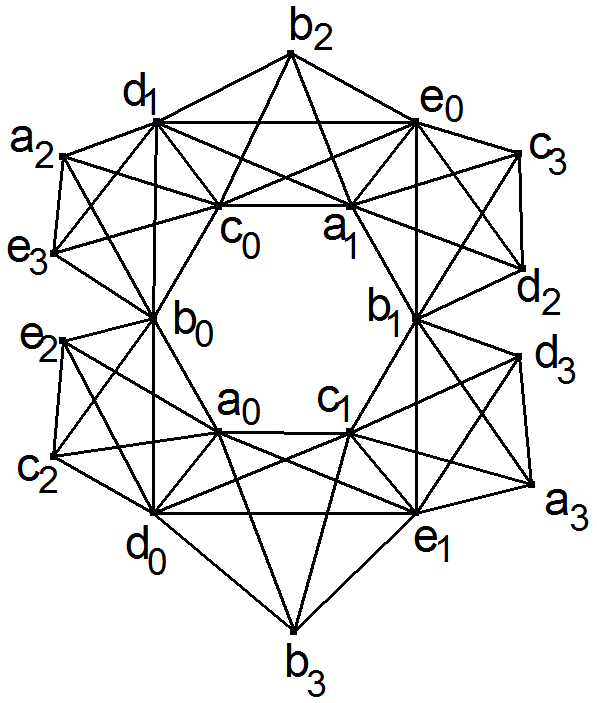}
\caption{Graph $G$}
\label{fig:abcde}
\end{center}
\end{minipage}\hfill
\begin{minipage}[b]{90mm}
\begin{center}
\includegraphics[scale=.75]{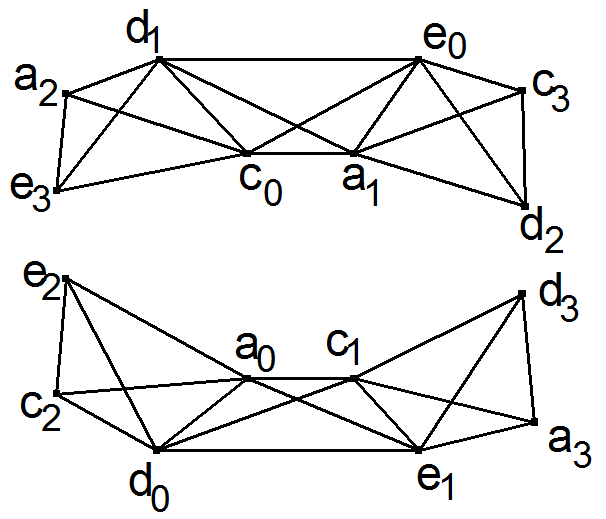}
\caption{The induced subgraph of $G$ on $4$ colours $a,c,d,e$. This graph is chordal.}
\label{fig:abde}
\end{center}
\end{minipage}
\end{figure}

Let $\mathcal{C}$ be the following set of characters:

$a  =  \{x,u\}|\{z,t\}|\{y\}|\{v\} = a_0|a_1|a_2|a_3$

$b  =  \{x,y\}|\{t,v\}|\{z\}|\{u\} = b_0|b_1|b_2|b_3$

$c = \{y,z\}|\{u,v\}|\{x\}|\{t\} = c_0|c_1|c_2|c_3 $

$d  =  \{x,u\}|\{y,z\}|\{t\}|\{v\} = d_0|d_1|d_2|d_3$

$e  = \{z,t\}|\{u,v\}|\{x\}|\{y\} = e_0|e_1|e_2|e_3$

Each character has $4$ states that we denote for example by $a_0,a_1,a_2,a_3$ for the character $a$. The partition intersection graph $G$ associated to $\mathcal{C}$ is in Figure \ref{fig:abcde}.

\begin{figure}[ht]
\begin{minipage}[b]{80mm}
\begin{center}
\includegraphics[scale=.75]{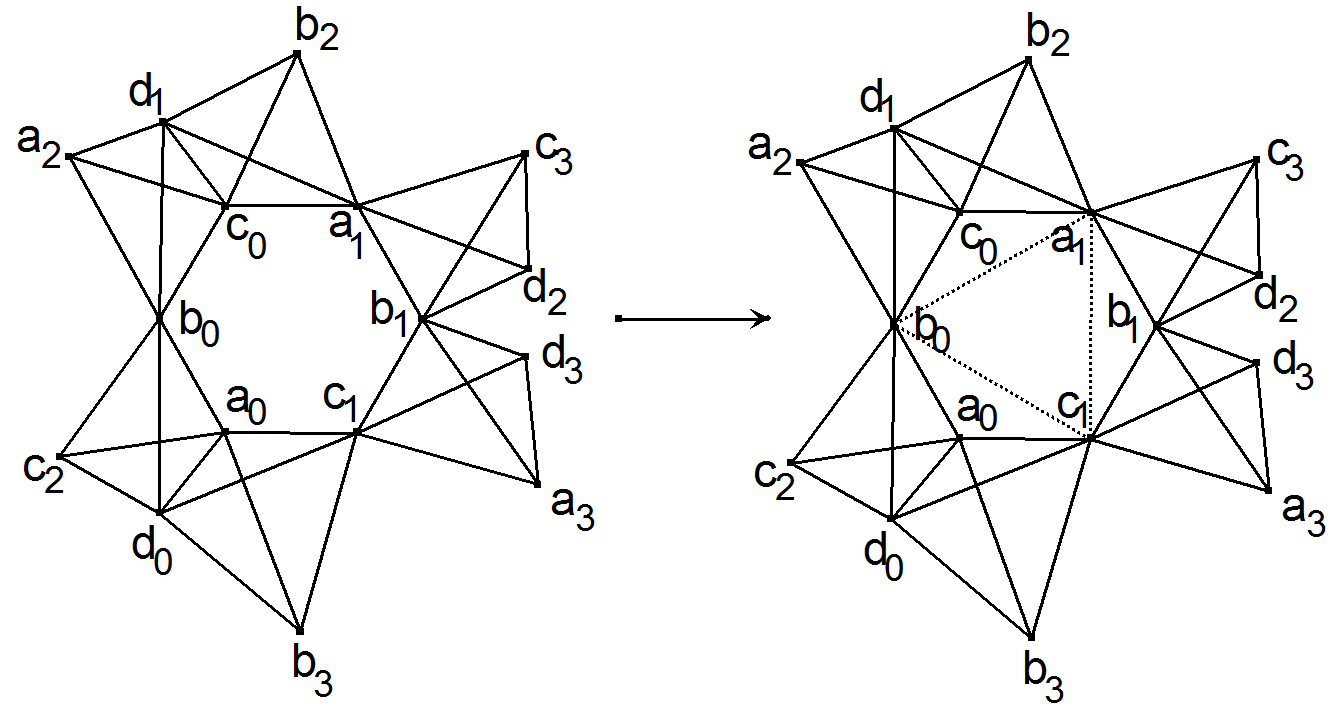}
\caption{The induced subgraph of $G$ on $4$ colours $a,b,c,d$ and its chordal completion}
\label{fig:abcd}
\end{center}
\end{minipage}\hfill
\begin{minipage}[b]{80mm}
\begin{center}
\includegraphics[scale=.75]{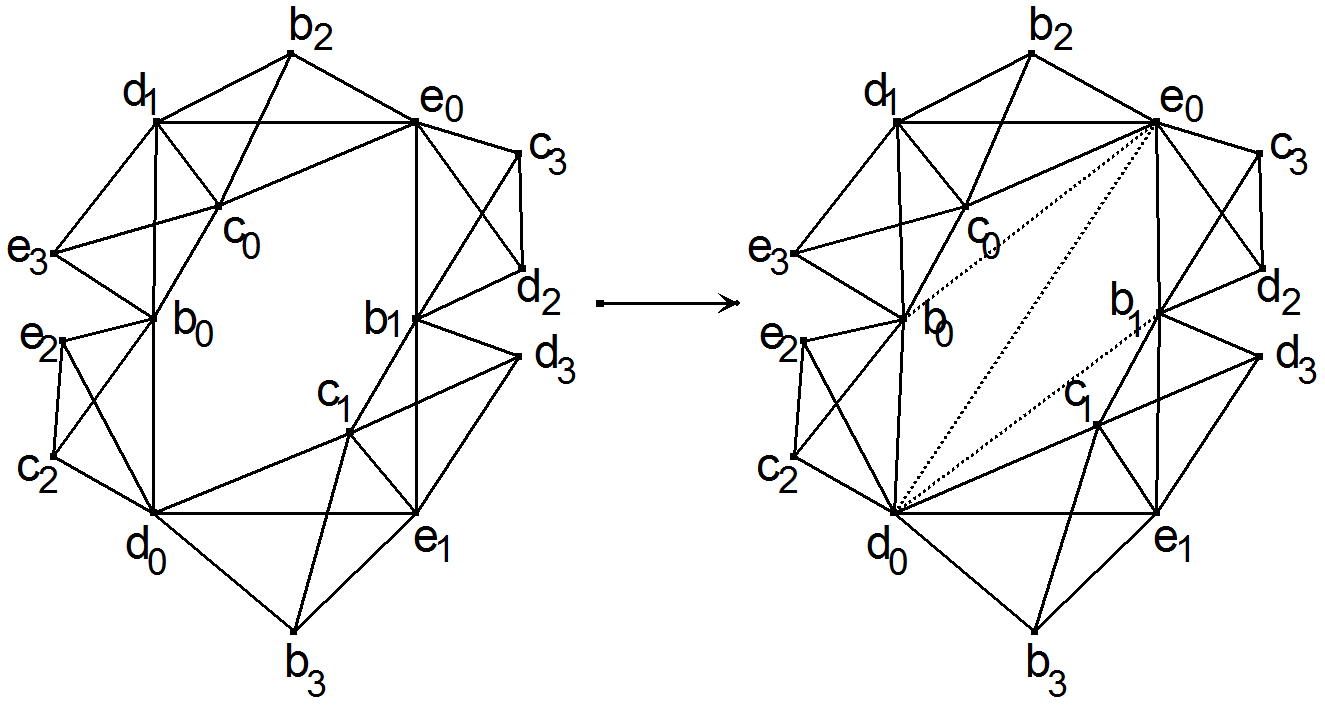}
\caption{The induced subgraph of $G$ on $4$ colours $b,c,d,e$ and its chordal completion}
\label{fig:bcde}
\end{center}
\end{minipage}
\end{figure}

$G$ does not accept any proper chordal completion. Indeed, if we consider only the induced subgraph of $G$ on $4$ colours $a,b,c,d$ and triangulate it, then there is a unique way to do that by connecting $(a_1,b_0), (b_0,c_1)$ and $(c_1,a_1)$ (Figure \ref{fig:abcd}). Similarly, if we consider the induced subgraph of $G$ on $4$ colours $a,b,c,e$, there is also a unique way to triangulate it by connecting $(a_0,b_1),(b_1,c_0)$ and $(c_0,a_0)$. 
So, to triangulate $G$, the cycle $(a_0b_1a_1b_0,a_0)$ is forced to be created. However, this cycle does not have any proper chordal completion. In other words, $G$ does not have any proper chordal completion. So, $\mathcal{C}$ is not compatible.

However, as we see in Figure  \ref{fig:abde}, the induced subgraph of $G$ on $4$ colours $a,c,d,e$ is chordal.
The induced subgraph of $G$ on $4$ colours $a,b,c,d$ has a proper chordal completion (Figure \ref{fig:abcd}) and similarly for the induced subgraph of $G$ on $4$ colours $a,b,c,e$ due to the symmetry. The induced subgraph of $G$ on $4$ colours $b,c,d,e$ also has a proper chordal completion (Figure \ref{fig:bcde}) and similarly for the induced subgraph of $G$ on $4$ colours $a,b,d,e$. 

It means that every $4$ characters of $\mathcal{C}$ are compatible but the whole set $\mathcal{C}$ is not compatible.


\section{A closure chordal sandwich graph operation}
\label{sec:partial}
Given a graph $G=(V,E,F)$ where $E \cap F = \emptyset$. Let $u,v$ be two vertices of $G$ such that $(u,v) \not\in E$. 

$(u,v)$ is a \textit{forbidden edge} if it is not included in any minimal proper chordal completion of $G$. So $(u,v)$ is forbidden if either $(u,v) \in F$ or if by connecting them, the resulting graph does not have any proper chordal completion. The forbidden edges are presented by dotted lines in our figures.

$(u,v)$ is a \textit{forced edge} if it is contained in every proper chordal completion of $G$. So if there is a cycle in $G$ which has a unique proper chordal completion, then the edges used to complete this cycle are forced. 

Note that by adding any forced edge into $E$ or any forbidden edge into $F$, we do not lose any proper chordal completion. 

A cycle $C$ of $G$ is forbidden if every chordal completion of $C$ contains at least an edge in $F$.
So if $G$ has a proper chordal completion then it does not have any forbidden cycle. The converse is not always true. For example, see the graph $G=(V,E,F)$ in Figure \ref{fig:ex_triangulation} where $F$ consists of the pairs of vertices having a same colour. This graph has $3$ chordless cycles and each one can be chordally completed without using any edges in $F$. However, $G$ does not admit any proper chordal completion.

\begin{example}
In Figure \ref{fig:ex1} we have a cycle of size $5$ on $3$ colours $a,b,c$. The set $F$ consists of $(a_0,a_1)$ and $(b_0,b_1)$. One can deduce that $(a_1,b_1)$ is forbidden since by connecting them we have the forbidden cycle $(a_0b_0a_1b_1)$. We deduce furthermore that $(c_0,a_0)$ and $(c_0,b_0)$ are forced because the unique way to properly chordally complete this cycle is  connecting them.

\begin{figure}[ht]
\begin{minipage}[b]{90mm}
\begin{center}
\includegraphics[scale=.7]{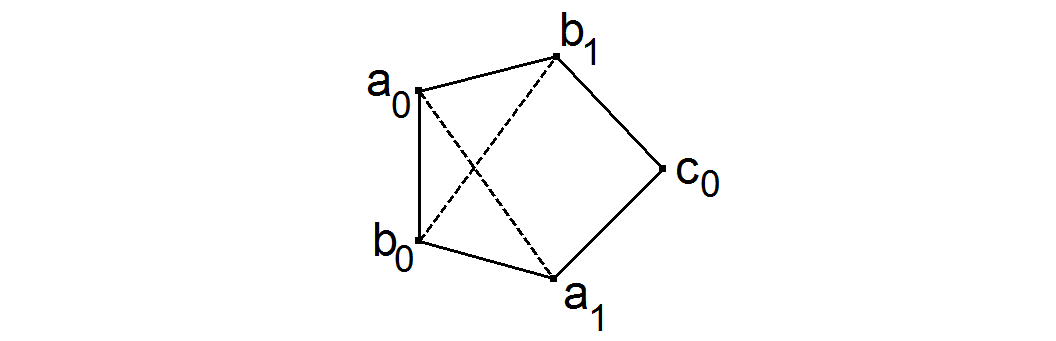}
\caption{$(a_0,a_1), (b_0,b_1), (a_1,b_1)$ are forbidden edges. $(c_0,a_0)$, $(c_0,b_0)$ are forced edges.}
\label{fig:ex1}
\end{center}
\end{minipage}\hfill
\begin{minipage}[b]{70mm}
\begin{center}
\includegraphics[scale=.7]{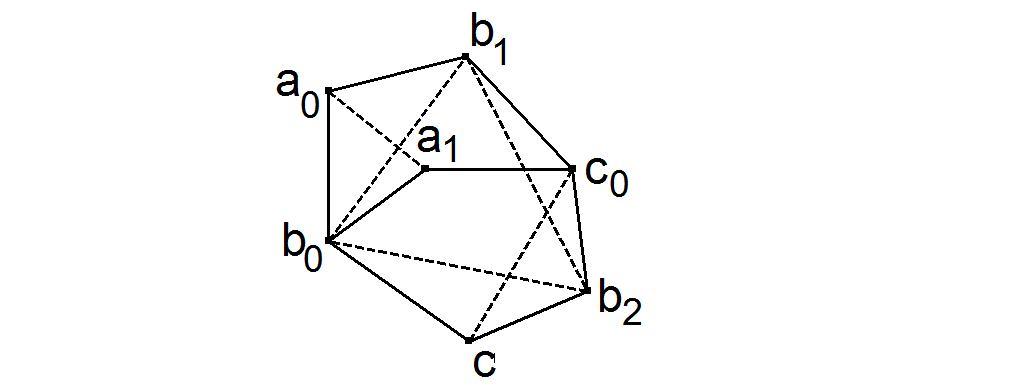}
\caption{A graph which does not have any proper chordal completion}
\label{fig:ex_triangulation}
\end{center}
\end{minipage}
\end{figure}

Consider the graph in Figure \ref{fig:ex_triangulation} where $F$ is the set of pairs of vertices having a same colour. Similarly to the previous example on the cycle $a_0b_1c_0a_1b_0$, we deduce that $c_0a_0$ and $c_0b_0$ are forced. So, the cycle $b_0c_0b_2c_1$ is forced to be presented in any proper chordal completion of this graph. However, this cycle is forbidden. Therefore, this graph does not have any proper chordal completion.
\end{example}

\begin{observation}Given a cycle $C=(u_1\dots u_k,u_1)$, then:

\label{ob:triangulate}
(i) for any $i \in \{1,\dots, k\}$, every chordal completion of $C$ must contain either $(u_{i-1},u_{i+1})$ or $(u_i,u_j)$ for a certain $j$ different from $i,i-1,i+1$. 

(ii) every chordal completion of $C$ must contain a chord $(u_{i-1},u_{i+1})$ for a certain $i$.
\end{observation}

\begin{lemma}[Detecting forbidden edges and forced edges] 

Let $G=(V,E,F)$ be a graph where $E \cap F = \emptyset$ and $(u,v)$ be two vertices of $G$:

\label{lem:forbidden_forced}
1) If there is a chordless path $(ut_1\dots t_kv)$ such that for any $i = 1, \dots k$, either $(u,t_i)$ or $(v,t_i)$ is forbidden, then $(u,v)$ is also forbidden. 

2) Suppose that $(u,v) \in E$. If there is a chordless cycle $c=(uwt_1\dots t_kv,u)$ such that for any $i=1, \dots k$, either $(u,t_i)$ or $(v,t_i)$ is forbidden, then $(v,w)$ is a forced edge.
\end{lemma}

\begin{figure}[ht]
\begin{minipage}[b]{70mm}
\begin{center}
\subfigure[$(u,v)$ is forbidden\label{fig:lemma_forbidden}]{\includegraphics[scale=.7]{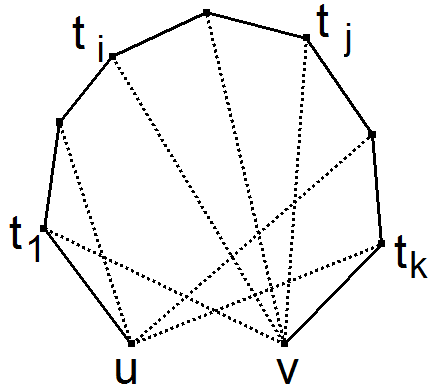}}\;\;\;\;\;\;
\subfigure[$(v,w)$ is a forced edge\label{fig:lemma_forced}]{\includegraphics[scale=.7]{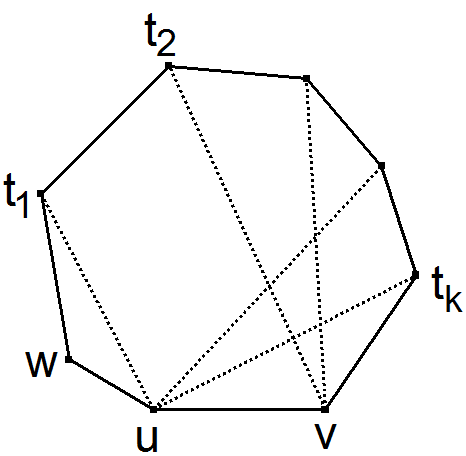}}
\caption{Lemma \ref{lem:forbidden_forced}}
\label{fig:forbidden_forced}
\end{center}
\end{minipage}\hfill
\begin{minipage}[b]{90mm}
\begin{center}
\subfigure[a $f(u,v)$ path, $(u,v)$ is forbidden.\label{fig:coro_forbidden}]{\includegraphics[scale=.7]{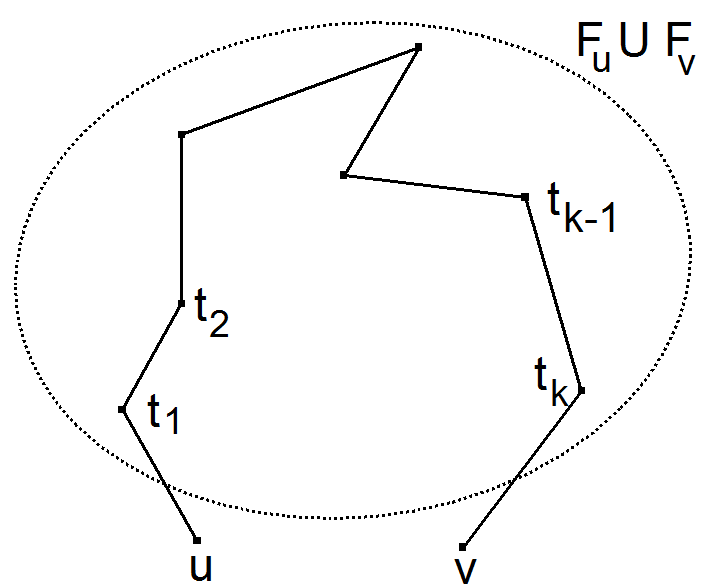}}
\subfigure[a $g(u,v,w)$ cycle. $(w,v)$ is forced.\label{fig:coro_forced}]{\includegraphics[scale=.7]{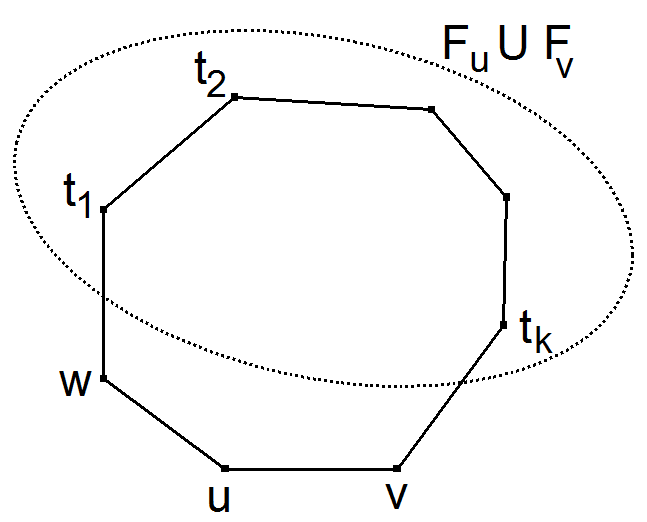}}
\caption{Corollary \ref{coro:forbidden_forced}}
\label{fig:forbidden_forced1}
\end{center}
\end{minipage}
\end{figure}

\begin{pf}
1) By connecting $(u,v)$, we obtain the chordless cycle $C=(ut_1\dots t_kv,u)$. We will prove the cycle $C$ is forbidden. By Observation \ref{ob:triangulate} (i), to chordally complete $C$, we must connect either $(t_1,v)$ or $(u,t_i)$ for a certain $i = 2,\dots,k$. However, $(t_1,v)$ is forbidden because $(u,t_1)$ is an edge of $G$ and by the assumption, either $(u,t_1)$ or $(v,t_1)$ must be forbidden.  So, we must connect an edge $(u,t_i)$, which must not  be a forbidden edge. We deduce that $(v,t_i)$ is forbidden. 
The created subcycle $(uu_iu_{i+1}\dots u_kv,u)$ has the same property as $C$. So, by using the same argument, to chordally complete this cycle, we must connect an edge $(u,u_j)$ where $i<j\le k$. So, the size of the considering cycle decreases strictly each time applying this argument. Then there will be a moment that we obtain a cycle which is forbidden, in other words the cycle $C$ is forbidden.

2) We will prove that, every proper chordal completion of $C$ must contain the chord $(v,w)$. Suppose that there is a proper chordal completion of this cycle which does not contain $(v,w)$. By Observation \ref{ob:triangulate} (i), this completion must contain $(u,t_i)$ for a certain $i=1,\dots,k$. We obtain then a subcycle $(ut_it_{i+1}\dots t_kv,u)$. The path $(ut_it_{i+1}\dots t_kv)$ satisfies the condition of Claim $1$, so $(u,v)$ is a forbidden edge. However, $(u,v)$ is an edge of $G$. It means that this subcycle does not have any proper chordal completion. In other words, $C$ does not have any proper chordal completion which does not contain $(u,v)$. So $(v,w)$ is presented in any chordal completion of $G$, i.e. it is a forced edge.
\end{pf}

\begin{corollary} Let a graph $G=(V,E,F)$, denote by $F(u)$ the set of vertices $u'$ such that $(u,u') \in F$. Then, for any pair of vertices $(u,v)$:

1) If $(u,v)$ is not an edge of $G$ and there is a chordless path $(ut_1\dots t_kv)$ such that for any $i = 1, \dots , k$, $t_i \in F(u) \cup F(v)$ 
then $(u,v)$ is forbidden. We call such a path a $f(u,v)$ path.

2) If $(u,v)$ is an edge of $G$ and there is a chordless cycle $(uwt_1\dots t_kv,u)$ such that $w \not\in F(v)$ and for any $i = 1, \dots, k$, $t_i \in F_u \cup F_v$, then $(v,w)$ is a forced edge. We call such a cycle a $g(u,v,w)$ cycle. 
\label{coro:forbidden_forced}
\end{corollary}

Denote by $N(u)$ the set of neighbour vertices of $u$.

\restylealgo{boxed}\linesnumbered
\begin{algorithm}[H]
\caption{A closure chordal-sandwich graph operation}
\label{algo:0}
\KwData{A graph $G=(V,E,F)$}
\KwResult{$Closure(G)$}

For any $u \in V$, calculate $N(u)$ and $F(u) = \{v|~(u,v) \in F\}$\;

flag = true\;

\While {(flag)}{
	flag = false\;
	\For {(any pair of vertices $(u,v)$)}{
		\If {(there is a $f(u,v)$ path)}{
			\If {($u \not\in N(v)$)}{				
				Add $u$ to $F(v)$, and $v$ to $F(u)$\;
				flag=true\;
			}
			\Else{
				$G$ does not have any proper chordal completion. \textit{Exit}\;
			}
		}
		\If{($u \in N(v)$) $\wedge$ (there is a $g(u,v,w)$ cycle)}{		
			Add $v$ to $N(w)$, and $w$ to $N(v)$\;
			flag=true\;		
		}
	}	
}
\Return $G'=(V,E',F')$ where $E'=\{(u,v)|~u\in N(v)\}$ and $F' = \{(u,v)|~u\in F(v)\}$.
\end{algorithm}

\begin{theorem}
Algorithm \ref{algo:0} takes time $O(n^4(n+m))$. Let $G'=(V,E',F') = Closure(G)$, then any proper chordal completion of $G'$ is a proper chordal completion of $G$ and vice-versa. Moreover, $G'$ satisfies the following properties:

1. For any $(u,v) \not\in E' \cup F'$, if we connect $(u,v)$ then for any created chordless cycle $C$ which has $(u,v)$ as an edge, $C$ has at least a chordal completion without using any pair of vertices in $F'$.

2. Any chordless cycle of $G'$ has at least two chordal completions without using any pair of vertices in $F'$.
\label{theo:partial}
\end{theorem}

\begin{pf}
\subsubsection*{Correctness:}

According to Corollary \ref{coro:forbidden_forced}, the loop \textit{for} recognizes  all pairs of vertices $(u,v)$ which satisfy the conditions in the corollary to detect the forbidden edges and forced edges. 
So the pairs of vertices added in $F'$ at line $8$ are the forbidden edges, i.e. the edges which are not included in any proper chordal completion of $G$. 
And the pairs of vertices added in $E'$ at line $13$ are forced edges, i.e. the edges which are included in every proper chordal completion of $G$. Therefore, any proper chordal completion of the obtained graph is a proper chordal completion of $G$ and vice versa. 

Moreover, if $(u,v)$ is an edge and there exists a $f(u,v)$ path then according to the proof of Lemma \ref{lem:forbidden_forced}, the chordless cycle consisting of $f(u,v)$ and $(u,v)$ is forbidden. Hence, $G$ does not have any proper chordal completion (line $11$). This process stops when there is no more forbidden edges or forced edges  detected. 
So, in $G'$, for any pair of vertices $(u,v)$, there is no $f(u,v)$ path;  and if $(u,v) \in E'$, then there is no $g(u,v,w)$ cycle. We will prove that $G'$ satisfies the two properties of the theorem.

\begin{figure}[ht]
\begin{center}
\subfigure[\label{fig:proof_algo1}]{\includegraphics[scale=.7]{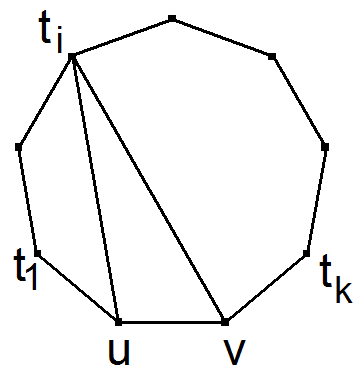}}\;\;\;\;\;
\subfigure[\label{fig:proof_algo3}]{\includegraphics[scale=.7]{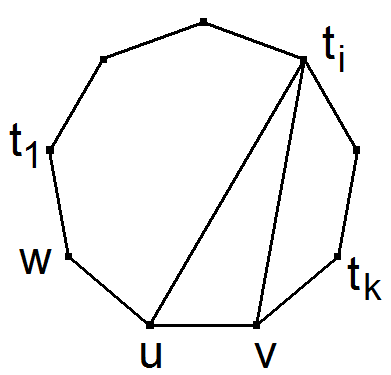}}
\caption{Proof of Theorem \ref{theo:partial}}
\end{center}
\end{figure}

1) The first property: we prove by induction on the size of cycles. 

Let $(u,v)$ be a pair of vertices not in $E' \cup F'$. By connecting $(u,v)$, let $C$ be a created cycle which contains $(u,v)$. We will prove that $C$ admits at least a chordal completion without using any edge in $F'$.

For the case $|C|=4$, let $C=(uxyv,u)$. By the assumption, $G'$ does not contain any $f(u,v)$ path, i.e. $(uxyv)$ is not a $f(u,v)$ path. It means that either $(v,x) \not\in F'$ or 
$(u,y) \not\in F'$. So, we can complete $C$ by connecting either $(v,x)$ or $(u,y)$.

Suppose that $C$ has at least a chordal completion if $|C| \le k$. 

For the case $|C|=k+1$, let $C = (ut_1\dots t_{k-1}v,u)$, so there is not any $f(u,v)$ path in $G'$. Then, there exists at least an $i \in \{1,\dots,k-1\}$ such that $(u,t_i),(v,t_i) \not\in F'$.
By the induction hypothesis, if we connect $(u,t_i)$, $(v,t_i)$, the two subcycles $(ut_1\dots t_i,u)$ and $(vt_i\dots t_k,v)$ have proper chordal completions without using any pair of vertices in $F'$ because both these two cycles have size smaller than $k+1$ (Figure \ref{fig:proof_algo1}). Completing these two subcycles gives a proper chordal completion for $C$. So, $C$ admits at least one chordal completion, the one which contains $(u,t_i)$ and $(v,t_i)$.

2) The second property: Suppose that there is a chordless cycle $C$ which admits a chordal completion without connecting any pair of vertices in $F'$.
By Observation \ref{ob:triangulate} (ii), this chordal completion must contain at least a triangle $(u,v,w)$ such that $(u,v)$, $(u,w)$ are edges of $C$.
Let $C=(uwt_1\dots t_kv,u)$, so $C$ is not a $g(u,v,w)$ cycle because otherwise $(v,w)$ is a forced edge and it must have been connected by the algorithm, a contradiction with the fact that $C$ is chordless. So, there is a $t_i$ such that $(u,t_i),(v,t_i) \not\in F'$ (Figure \ref{fig:proof_algo3}). Using the first property, if we connect $(u,t_i)$ and $(v,t_i)$, then we obtain two subcycles which have proper chordal completions without connecting any pair of vertices in $F'$. That implies another chordal completion of $C$ containing $(u,t_i)$ and $(v,t_i)$. This chordal completion does not contain $(v,w)$, so it is different with the initial one. In other words, $C$ has at least $2$ distinct proper chordal completions without using any pair of vertices in $F'$.

\subsubsection*{Complexity:}

\hspace*{.5cm}- Calculating $N(u)$ and $F(u)$ for any vertex $u$ in line $1$ is done in times $O(n^2)$.

- The loop \textit{while}: For each iteration, there is at least a pair of vertices $(u,v)$ whose nature is modified, i.e $(u,v)$ becomes either a forbidden edge or a forced edge. Once it is modified, it will not be modified afterwards. The loop stops when there is no more modification on any pair of vertices. So, the number of iterations of this loop is bounded by the number of pairs of vertices, i.e by $O(n^2)$.

- The loop \textit{for}: there are $O(n^2)$ pairs of vertices $(u,v)$. So there are $O(n^2)$ iterations. In each iteration:

\hspace*{.6cm} $\bullet$ Checking if there is a simple path $f(u,v)$ can be done in linear time: We proceed a dfs starting at $u$ such that the visited vertices are in $F(u) \cup F(v) \setminus N(u)$. If  we meet a vertex in $N(v)$, then there is a $f(u,v)$ path. Otherwise, there is no such path.

\hspace*{.5cm} $\bullet$ Checking if there is a $g(u,v,w)$ cycle can also be done in linear time: We proceed a dfs from $u$ such that the first visited vertices is not in $N(v) \cup F(v)$, and the remaining visited vertices are in $F(u) \cup F(v) \setminus N(u)$. If we meet a vertex in $N(v)$ then we have a $g(u,v,w)$ cycle. Otherwise, there is no such cycle. 

So, the total complexity is $O(n^4(n+m))$ where $n$ is the number of vertices of $G'$ and $m$ is the number of edges of the obtained graph.
\end{pf}

With this operation, one can deduce for example that the input graph does not contain any cycle as in Figure \ref{fig:ex1} because this cycle must be already triangulate by the operation. Or the input graph can not contain an induced subgraph as in Figure \ref{fig:ex_triangulation}, because the operation can show that in this case the graph does not have any proper chordal completion. 

\begin{corollary}
Without loss of generality, one can suppose that any cycle of a graph $G=(V,E,F)$ has at least two proper chordal completions; and by connecting any pair of vertices not in $E \cup F$, every created cycle has at least one proper chordal completion.
\end{corollary}


\section{Conclusion}
\label{sec:conclusion}
Our example in Section \ref{sec:example} showed that $f(4) \ge 5$. We suggest that $f(r) \ge r+1$ for any $r \ge 4$. So, a further work is to generalize our example, or to find other examples supporting this suggestion. Another problem is proving that $f(r)$ exists by determining an upper bound function of $r$ for $f(r)$. It means that we must find a function $F(r)$ such that if every set $F(r)$ characters of $\mathcal{C}$ is compatible then $\mathcal{C}$ is compatible. A harder question is determining $f(r)$ for $r \ge 4$. Our closure operation for chordal sandwich graphs can help to simplify these problems. 

\bibliography{Thesis}
\bibliographystyle{plain}

\end{document}